\documentclass{ws-ijmpc}

\usepackage{amssymb}
\usepackage{comment}
\usepackage{eurosym}   
\usepackage{mathrsfs}  

\DeclareRobustCommand{\Eqref}[1]{Eq.~(\ref{#1})}

\usepackage{booktabs} 
\usepackage{dcolumn} 
\newcolumntype{d}[1]{D{.}{.}{#1}}
\usepackage{multirow}

\usepackage{todo}     

\usepackage{url}      

\usepackage{courier}

\usepackage{float}
\usepackage{listings} 
\usepackage{lineno}

\usepackage{siunitx}
\usepackage{algorithmic}
\usepackage{algorithm}
\usepackage[normalem]{ulem}

\begin{document}


\markboth{C. Bonati \& E. Calore \& S. Coscetti \& M. D'Elia \& M. Mesiti \& 
  F. Negro \& S.F. Schifano \& G. Silvi \& R. Tripiccione}
{Design and optimization of portable LQCD Monte Carlo code using OpenACC}

\catchline{}{}{}{}{}

\title{Design and optimization of a portable LQCD Monte Carlo 
code using OpenACC}

\author{Claudio Bonati\footnote{Present address: Dipartimento di Fisica e 
  Astronomia and INFN-Sezione di Firenze, 
Via Sansone 1, 50019 Sesto Fiorentino (FI), Italy.} }
\address{
Universit\`a di Pisa and INFN Sezione di Pisa,\\
Largo Pontecorvo 3, I-56127 Pisa, Italy\\
claudio.bonati@df.unipi.it
}
\author{Enrico Calore}
\address{Universit\`a degli Studi di Ferrara and INFN Sezione di Ferrara,\\
Via Saragat 1, I-44122 Ferrara, Italy\\
enrico.calore@fe.infn.it}
\author{Simone Coscetti}
\address{INFN Sezione di Pisa,\\ 
Largo Pontecorvo 3, I-56127 Pisa, Italy\\
simone.coscetti@pi.infn.it
}
\author{Massimo D'Elia}
\address{Universit\`a di Pisa and INFN Sezione di Pisa,\\ 
Largo Pontecorvo 3, I-56127 Pisa, Italy\\
massimo.delia@unipi.it
}
\author{Michele Mesiti}
\address{Universit\`a di Pisa and INFN Sezione di Pisa,\\
Largo Pontecorvo 3, I-56127 Pisa, Italy\\
michele.mesiti@pi.infn.it
}
\author{Francesco Negro}
\address{INFN Sezione di Pisa,\\ 
Largo Pontecorvo 3, I-56127 Pisa, Italy\\
fnegro@pi.infn.it
}
\author{Sebastiano Fabio Schifano}
\address{Universit\`a degli Studi di Ferrara and INFN Sezione di Ferrara,\\ 
Via Saragat 1, I-44122 Ferrara, Italy\\
schifano@fe.infn.it
}
\author{Giorgio Silvi\footnote{Present address: Juelich Supercomputing Centre, 
Forschungszentrum Juelich, 52428 Juelich, Germany (g.silvi@fz-juelich.de).}}
\address{Universit\`a degli Studi di Ferrara and INFN Sezione di Ferrara,\\ 
Via Saragat 1, I-44122 Ferrara, Italy\\
giorgiosilvi@gmail.com
}
\author{Raffaele Tripiccione}
\address{Universit\`a degli Studi di Ferrara and INFN Sezione di Ferrara,\\
Via Saragat 1, I-44122 Ferrara, Italy\\
tripiccione@fe.infn.it
}

\maketitle

\begin{history}
\received{Day Month Year}
\revised{Day Month Year}
\end{history}


\begin{abstract}


The present panorama of HPC architectures is extremely heterogeneous, ranging
from traditional multi-core CPU processors, supporting a wide class of
applications but delivering moderate computing performance, to many-core GPUs,
exploiting aggressive data-parallelism and delivering higher performances for
streaming computing applications.
In this scenario, code portability (and performance portability) become
necessary for easy maintainability of applications; this is very relevant in
scientific computing where code changes are very frequent, making it tedious
and prone to error to keep different code versions aligned.
In this work we present the design and optimization of a state-of-the-art
production-level LQCD Monte Carlo application, using the directive-based
OpenACC programming model. OpenACC abstracts parallel programming to a
descriptive level, relieving programmers from specifying how codes should be
mapped onto the target architecture. 
We describe the implementation of a code fully written in OpenACC, and show
that we are able to target several different architectures, including
state-of-the-art traditional CPUs and GPUs, with the same code.
We also measure performance, evaluating the computing efficiency of our OpenACC
code on several architectures, comparing with GPU-specific implementations and
showing that a good level of performance-portability can be reached. 

\keywords{LQCD; Portability; OpenACC; Graphics Processing Units; GPU}

\end{abstract}

\ccode{PACS Nos.: 
07.05.Bx 
12.38.Gc 
}


\section{Introduction and related works}

The use of processors based on multi- and many-core architectures is a common
option in High Performance Computing (HPC).
Several variants of these processors exist, differing mainly in the number and
architecture of the cores integrated in a single silicon die.

Conventional CPUs integrate tens of fat cores sharing a large on-chip cache.
Fat cores include several levels of caches and complex control structures, able
to perform hardware optimization techniques (branch-speculation, instruction
scheduling, register renaming, etc).  Vector instructions are also supported by
these cores, with a moderate level of data parallelism: 2 to 4 vector elements
are processed by one vector instruction.
This architecture is reasonably efficient for many type of regular 
and non-regular applications and delivers a level of
performance of the order of hundreds of GigaFlops per processor.

On the other side of the spectrum we have Graphics Processor Units (GPU),
available as accelerator boards attached to conventional CPUs. 
GPUs integrate thousands of slim cores able to efficiently support regular
streams of computation, and deliver performances of the order of several
TeraFlops.  GPUs are extremely aggressive in terms of data-parallelism,
implementing vector units with large vector sizes (16 and 32 words are
presently available options).  

Midway between these two architectures, we have the Intel {\em Many 
Integrated Cores} (MIC) architecture based on several tens of slim cores. 
In this case, cores are similar to their fat counterparts, but their design has 
been simplified removing many hardware control  structures (instruction 
scheduler, register renaming, etc) and adopting wider vector units, able to 
process up to 4 or 8 vector elements in parallel. 

Large scale computing centers today have not reached a common consensus on the
``best'' processor option for HPC systems, also because system choices are
driven not only by application performances, but also by cost of ownership and
energy aspects which are becoming increasingly critical
parameters\cite{villa14}. Several computing centers do adopt machines based on
GPUs, but other ones prefer to stay on more traditional CPUs, offering a lower
peak performance, but better computing efficiency for a wider range of
applications.

In this scenario, the development of applications would greatly benefit from
the availability of a unique code version, written in an appropriate
programming framework, able to offer portability, in terms of code and
performance, across several present and possibly future state-of-the-art
processor architectures. 
A single code version, portable across several architectures, is of great
convenience in particular for scientific applications, where code changes and
development iterations are very frequent, so keeping several
architecture-specific  code  versions up-to-date is a tedious and  error prone
effort\cite{se4hpcs15,scaleff-maintport-review}.

Directives based programming models are going exactly in this direction, 
abstracting parallel programming to a descriptive level as opposite to a 
prescriptive level, where programmers must specify how the code should be mapped
onto the target machine.
OpenMP\cite{openmp} and OpenACC\cite{openacc} are among the most common such
programming models, already used by a wide scientific community. 
Both are based on directives: OpenMP was introduced to manage parallelism on
traditional multi-core CPUs, while OpenACC is mainly used to target GPUs
(although designed to be architecture agnostic)\cite{Wienke2014812}.
These two frameworks are in fact converging and extending their scope to cover
a large subset of HPC applications and architectures: OpenMP version 4 has been
designed to support also accelerators, while compilers supporting OpenACC (such
as the PGI\cite{PGIref}) are starting to use directives also to target
multi-core CPUs.

In this work we describe the implementation of a Lattice QCD (LQCD) Monte Carlo 
code designed to be portable and efficient across several architectures.
LQCD simulations represent a typical and well known HPC grand challenge, with
physics results strongly limited by available computational
resources\cite{Bernard:2002pd,bilardi}; over the years, several generations of
parallel machines, optimized for LQCD, have been
developed\cite{ape,apecise,qcdoc,qpacepower,bgq}, while the development of LQCD
codes running on many core architectures, in particular GPUs, has seen large
efforts in the last 10
years\cite{videogame,Barros:2008rd,quda,Cardoso:2010di,Chiu:2011dz,cudacode,bach1,bach2}.
Our goal is to have just one code able to run on several processors without any
major code changes, and possibly to have roughly the same level of efficiency,
looking for an acceptable trade-off between portability and
efficiency\cite{scaleff-maintport-review,perfport-directives}.
%

As a programming model we have selected OpenACC, as it currently 
has a wider compiler support, in particular targeting NVIDIA GPUs, which 
are widely used in HPC clusters and commonly used for scientific computations. 
OpenACC has been successfully used to port and run other scientific codes, such 
as Lattice Boltzmann applications\cite{blair15,jiri,ccpe16} in computational 
fluid-dynamics, showing a good level of code and performance portability on 
several architectures. 
The migration of our code to OpenMP4, if needed, as soon as compiler support 
becomes more mature, is expected to be a simple additional effort.

We have developed a code with all key features for a 
state-of-the-art simulations of QCD with dynamical fermions. 
Using this code as a user test case, we  assess: 
i) if it is possible to write the code in such a way that the most 
computationally critical kernels can be executed on accelerators, 
as in previous CUDA implementations\cite{cudacode}; 
ii) how many of the presently available multi and many-core architectures can be
really used; 
iii) how efficient are these codes, and in particular what is the price to pay 
in terms of performance with respect to a code written and optimized for 
a specific architecture (e.g., using CUDA for GPUs).

We believe that our work is a non trivial step forward in the development of a fully 
portable production-grade LQCD Monte Carlo code, using the OpenACC programming model. 
An earlier paper\cite{pushan16} presented tests of selected portions of an 
OpenACC LQCD implementation on Fermi and K20 NVIDIA GPUs,
comparing performances with an OpenMP implementation for CPUs.
Similarly, in a preliminary study\cite{se4hpcs15}, we compared the 
performance of selected kernels of a full simulation, written in OpenACC, 
with an equivalent CUDA implementation, on a K20 NVIDIA GPU.
In this work, we extend the use of OpenACC in several new directions: 
i) we show the portability of a complete implementation across several 
architectures; 
ii) we show performance figures for the same OpenACC code on a variety of 
multi and many-core processors, including the most recent GPUs like the K80 and 
the recently released P100;
iii) we compare results with a previous implementation of the same full 
application written in CUDA\cite{cudacode}. 

The remainder of the paper is organized as follows: in Section~\ref{simalg} 
we give a brief introduction to LQCD and to the main computational
aspects of our application;
in Section~\ref{hpctrend} we highlight recent 
developments in HPC hardware and programming tools; in 
Section~\ref{implementation} 
we describe the OpenACC implementation of our code; in 
Section~\ref{results} we analyze 
our results; finally,
Section~\ref{conclusions}, contains our concluding remarks. 


\section{Numerical challenges of Lattice QCD}\label{simalg}


Quantum Chromodynamics (QCD) is the quantum field theory that describes strong
interactions in the Standard Model of particle physics. It is a non-abelian
gauge theory, based on the $SU(3)$ group (the ``color'' group), describing the
interactions of six different species (``flavors'') of quarks, mediated by $8$
vector bosons, the ``gluons''.

In principle QCD is not different from the theory that describes other sectors
of the Standard Model (i.e. the electroweak interaction); however, strong
interactions are indeed strong, i.e. the coupling constant of QCD is
generically not small. Asymptotic freedom ensures that the coupling constant
gets smaller and smaller as the energy scale increases (a summary of
experimental results is available in \S9.4 of the Particle Data Group
review\cite{pdg}), but a wealth of interesting phenomena take place for
energies well below the perturbative regime; a systematically improvable
computational scheme, that does not rely on the smallness of the coupling
constant, is needed to study this phenomenology from first principles. Lattice
QCD provides such a scheme. 

LQCD uses the Feynman path-integral quantization and approximates the infinite
dimensional path-integral by a finite dimensional integral: continuous
space-time is replaced by a finite lattice of sizes $L_t$, $L_x$, $L_y$, $L_z$
and lattice spacing $a$. In order to maintain gauge invariance, the variables
$U_{\mu}(n)$ associated with the gauge fields are elements of the $SU(3)$ group
and live on the links of the lattice; the quark fields $\psi(n)$ live on the
lattice sites and transform under the gauge group as $3-$dimensional complex
vectors\cite{Wilson:1974sk}. 
The fundamental problem of LQCD is the evaluation of expectation values of 
given functions of the fields, $O[U]$, that is integrals of the form   
\begin{eqnarray}\label{eq:pathint}
\langle \hat{O}\rangle =\frac{1}{Z}\int \mathscr{D}U O[U]\det(M[U])e^{-S_g[U]}\ , \quad
Z =\int \mathscr{D}U \det(M[U])e^{-S_g[U]}\ ;
\end{eqnarray}
the exponent $S_g$ is the discretization of the action of the gauge fields
(usually written as a sum of traces of products of $U_{\mu}(n)$ along closed
loops) and $\det(M)$ describes the gluon-quark interaction. Here, $M[U]$ is a
large and sparse structured matrix (i.e. containing both space-time and color
indexes) which is the discretization of the continuum fermion operator $M \sim
m\, {\rm I} + D$ where $m$ is the fermion mass, multiplying the identity
operator, and $D$ is the Dirac operator, which is constructed in terms of
covariant derivatives.  The integral in $\mathscr{D}U$ extends over all the
$U_{\mu}(n)$ variables on the lattice using the Haar measure of $SU(3)$.
Eq.~(\ref{eq:pathint}) refers to a single quark species (flavor); in the
realistic case of multiple flavors\footnote{At present, we have experimental
evidence of 6 different flavors in Nature, usually named with the letters $u$,
$d$, $s$, $c$, $b$, $t$ and ordered by increasing quark mass. In a realistic
simulation, one usually takes into account the first 3 (or 4, at most) flavors,
since the heaviest species give a negligible contribution to the low-energy
dynamics of the theory.}, one has to introduce a separate determinant for each
flavor.

This formulation makes contact with a standard problem in statistical
mechanics: importance sampling of the distribution $\det(M[U])e^{-S_g[U]}$.
What is non-standard is the form of this distribution and in particular the
presence of the determinant.  The best strategy devised so far to cope with
this problem is to introduce the so called pseudofermion
fields\cite{Weingarten:1980hx} $\phi$ and rewrite the integral as follows:
\begin{eqnarray}\label{eq:pseudofermions}
\int \mathscr{D}U O[U]\det(M[U]) e^{-S_g[U]}\propto 
\int \mathscr{D}U \mathscr{D}\phi \, O[U]\exp\left(-S_g[U]-\phi^{\dag} M[U]^{-1}\phi\right)\ ;
\end{eqnarray}
the action is still a non-local function of the field variables, but the
computational burden  required for the solution of a large sparse linear system
is much lower than the one needed for the computation of its determinant.

The explicit form of $S_g[U]$ and $M[U]$ is not fully determined, as these
functions only have the constraint to go over to the correct continuum limit as
the lattice spacing goes to zero. Much in the same way as several
discretization schemes exist for the numerical solution of a partial
differential equation, several discretization schemes of the QCD action exist.
In this paper we consider a specific state-of-the-art discretization, the
tree-level Symanzik improved action\cite{Weisz:1982zw,Curci:1983an} for the
gauge part and the stout-improved\cite{Morningstar:2003gk} ``staggered'' action
for the fermion part.  Staggered actions have a residual degeneracy, that has
to be removed by taking the $4-$th root of the determinant.  So,
\Eqref{eq:pseudofermions} becomes in the staggered case
\begin{eqnarray}\label{eq:rooting}
\int \mathscr{D}U \mathscr{D}\phi \, O[U]\exp\big(-S_g[U]-\phi^{\dag} M[U]^{-1/4}\phi\big)\ .
\end{eqnarray}

\subsection{Why LQCD is a computational grand challenge}
\label{grand_challenge_section}

The physical system that one would like to simulate by the lattice box has a
characteristic physical length $\xi$, which is of the order of $10^{-15}$ m. In
order to reduce systematic effects related to discretization and to the finite
box size, one would like that, at the same time, the lattice spacing $a$ be
much smaller, and the box size $L a$ much larger than $\xi$, i.e. $a \ll \xi
\ll La$.  Making the reasonable approximation that $\ll$ translates into one
order of magnitude means that the number of sites in each direction should be
$\simeq10^2$; the corresponding fermion matrix, considering also internal
(e.g., color) indexes, has a dimension slightly exceeding $10^8 \times 10^8$;
note that it is a sparse matrix, since the discretization of the Dirac operator $D$
connects only neighbor lattice sites.  
In finite temperature simulations the size of the lattice is typically smaller,
since in that case the temporal direction is shortened and equal to the inverse
of the temperature, $1/T$. 

The most computationally demanding task in the typical LQCD algorithm is the
solution of a linear system involving the fermion matrix $M$. The numerical
difficulty of this problem is fixed by the condition number of $M$, hence,
since the highest eigenvalue is typically $O(1)$, by the smallest eigenvalue of
$M$.  Here the physical properties of QCD play a significant role: the
eigenvalues of the Dirac operator are dense around zero, a property related to
the so-called {\em spontaneous breaking of chiral symmetry}, so the smallest
eigenvalue is set by $a m$ where $m$ is quark mass. Since Nature provides us
with two quark flavors ($u$ and $d$ quarks) whose mass is significantly lower
(by two orders of magnitude) than other energy scales of the theory, typical
values of $a m$ are typically very small, resulting in a bad condition number
($\kappa\gtrsim 10^5$ being a typical value).  Also regarding this aspect, the
situation becomes better when one is interested in the regime of very high
temperatures, since in that case the spontaneous breaking of chiral symmetry
disappears, the minimum eigenvalue of $D$ is non-zero, and the condition number
significantly improves.

\subsection{Numerical algorithms for LQCD}
\label{algorithm_section}
In LQCD, the usual local updates adopted in statistical mechanics scale badly
with the volume, as the action of \Eqref{eq:pseudofermions} is non-local. This
problem is partly solved by the Hybrid Monte Carlo (HMC)
algorithm\cite{Duane:1987de}; in HMC we associate fake conjugate momenta --
entering quadratically in the action --  to each degree of freedom of the
system. For an $SU(3)$ gauge theory, momenta conjugate to the link variable are
again $3 \times 3$ matrices $H_\mu(n)$ associated to each link of the lattice,
this time living in the group algebra (hence Hermitian and traceless).
Eq.~(\ref{eq:rooting}) is rewritten as  
\begin{eqnarray}\label{eq:rooting2}
\int \mathscr{D}U \mathscr{D}\phi 
\mathscr{D} H  
O[U]\exp\left(-\frac{1}{2} H^2 -S_g[U]-\phi^{\dag} M[U]^{-1/4}\phi\right)\ ,
\end{eqnarray}
where the momenta term is a shorthand to indicate the sum of $- {\rm Tr}
(H_\mu(n)^2)/2$ over the whole lattice.  The update then proceeds as follows:
\begin{enumerate}
\item random gaussian initial momenta $H$ and pseudofermions 
$\phi$ are generated;
\item starting from the initial configuration and momenta $(U,H)$, a new
state $(U', H')$ is generated by integrating the equations of motion;  
\item the new state $(U', H')$ is accepted with probability $e^{-\Delta S}$, where
$\Delta S$ is the change of the total (i.e. included the momenta) action.
\end{enumerate}
Step 2 is an unphysical evolution in a fictitious time and, under mild
conditions on the numerical integration of the equations of motion, it can be
shown to satisfy the detailed balance principle\cite{Duane:1987de,KennedyLec},
so it provides a stochastically exact way to estimate the integral in
\Eqref{eq:pseudofermions}.  The more time consuming steps of the update are the
ones that involve the non-local term in the exponent of
\Eqref{eq:pseudofermions}.  In particular, the most time consuming single step
of the whole algorithm is the solution of a linear system
\begin{equation}\label{eq:lineq}
M[U]\varphi=b \, .
\end{equation}
This calculation is needed to compute the forces appearing in the equations of
motion and also to evaluate $\Delta S$, and one usually resorts to Krylov
solvers.
In the case of staggered fermions, corresponding to \Eqref{eq:rooting}, it is
customary to use the so-called Rational HMC (RHMC)
algorithm\cite{Clark:2004cp,Clark:2006fx,Clark:2006wp}, in which the algebraic
matrix function appearing in \Eqref{eq:rooting} is approximated to machine
precision by a rational function. In this case one replaces \Eqref{eq:lineq} by
$r$ equations ($r$ is the order of the approximation adopted)
\begin{equation}\label{eq:shlineq}
(M[U]+\sigma_i)\varphi_i=b\ , \quad i\in\{1,\ldots,r\}\ , 
\end{equation}
where the real numbers $\sigma_i$ are the poles of the rational approximations.
These equations can again be solved by using Krylov methods: by exploiting the
shift-invariance of the Krylov subspace it is possible to write efficient
algorithms that solve 
all the equations appearing in (\ref{eq:shlineq}) at the same time, using
at each iteration only one matrix-vector product\cite{Jegerlehner:1996pm,Simoncini}.

For most of the discretizations adopted in QCD (and in particular for the one
we use), the matrix $M[U]$ can be written in block form
\begin{equation}\label{eq:diracblock}
M=m\,I+\left(\begin{array}{cc} 0 & D_{oe} \\ D_{eo} & 0
\end{array}\right), \qquad
D_{oe}^{\dag}=-D_{eo}\ ; 
\end{equation} 
matrices $D_{oe}$ and $D_{eo}$ connect only even and odd sites. It is
thus convenient to use an even/odd preconditioning\cite{DeGrandDeTarBook,DeGrand:1990dk}; 
in this case, \Eqref{eq:lineq} is replaced by:
\begin{equation}\label{eq:lineqeo}
(m^2\, I-D_{eo}D_{oe})\varphi_{e}=b_{e};
\end{equation}
$\varphi_e$ is defined only on even sites and the matrix is positive
definite (because of \Eqref{eq:diracblock}), so we can use the simplest of
the Krylov solvers: the conjugate gradient (or its shifted counterpart).

Over the years, many improvements of this basic scheme have been developed;
these are instrumental in reducing the computational cost of actual simulations
but their implementation is straightforward, once the basic steps of the
``naive'' code are ready. For this reason we will not discuss in the
following the details of multi-step integrators\cite{Sexton:1992nu,Urbach:2005ji}, 
improved integrators\cite{Omelyan02,Omelyan03,Takaishi:2005tz}, multiple 
pseudofermions\cite{Clark:2006fx} or the use
of different rational approximations and stopping residuals
in different parts of the HMC\cite{Clark:2006wp}, even if our code uses all these improvements.

\subsection{Data structures and computational challenges}

Our most important data structures are the collection of all gauge
variables $U_{\mu}(n)$ (elements of the group of $SU(3)$  matrices, one for each
link of the four-dimensional lattice) and of the pseudofermion fields $\phi(n)$
($3-$dimensional complex vectors, one for each even site of the lattice when
using the even/odd preconditioning).
We also need many derived and temporary data structures, such as:
\begin{enumerate}
\item the configurations corresponding to different stout levels
($U_{\mu}^{(k)}(n)$, again $SU(3)$ matrices), used in the computation of the
force (typically less than five stout levels are used) and the momenta
configuration (which are $3 \times 3$ Hermitian traceless matrices);
\item some auxiliary structures needed to
compute the force acting on the gauge variables, like the so called
``staples'' $\Sigma_{\mu}^{(k)}(n)$ and the $\Gamma_{\mu}(n)$ and
$\Lambda_{\mu}(n)$ matrices\cite{Morningstar:2003gk};
$\Sigma_{\mu}^{(k)}(n)$ and $\Gamma_{\mu}(n)$ are generic $3\times 3$ complex
matrices and $\Lambda_{\mu}(n)$ are $3\times 3$ Hermitian traceless matrices;  
\item the solutions $\varphi_i$ of \Eqref{eq:shlineq} and some auxiliary 
pseudofermion-like structure needed in the Krylov solver.
\end{enumerate}

At the lowest level, almost all functions repeatedly multiply two $3\times 3$
complex matrices (e.g., in the update of the gauge part), or a $3\times 3$
complex matrix and a $3-$dimensional complex vector (e.g., in the Krylov
solver) or compute dot products and linear combinations of complex $3-$vectors.
All these operations have low computational intensity, so it is convenient to
compress as much as possible all basic structures by exploiting their algebraic
properties.  The prototypical example is $U_{\mu}(n)$: one only stores the
first two rows of the matrix and recovers the third one on the fly as the
complex conjugate of the wedge product of the first two
rows\cite{DeForcrand:1986inu}.  This overhead is negligible with respect to the
gain induced, at least for GPUs, by the reduction of the memory
transfer\cite{Clark:2009wm,JooPhi}\footnote{A priori it would be possible to do
even better, i.e. to store just $8$ real numbers, but in this case the
reconstruction algorithm presents some instabilities\cite{Clark:2009wm}.}.

At a higher level the single most time consuming function is the Krylov solver,
which may take $40 \dots 80\%$ of the total execution time of a realistic
simulation (depending e.g. on the value of the temperature) and consists
basically of repeated applications\footnote{typically $10^2\div 10^3$
iterations are needed to reach convergence, depending on the temperature.}  of
the $D_{oe}$ and $D_{eo}$ matrices defined in \Eqref{eq:diracblock}, together
with some linear algebra on the pseudofermion vectors (basically
\emph{zaxpy}-like functions).  An efficient implementation of $D_{eo}$ and
$D_{oe}$ multiplies is then of paramount importance, the effectiveness  of this
operation being often taken as a key figure of merit in the LQCD community.


\section{Current trends in HPC}
\label{hpctrend}


There is a clear trend in high-performance computing (HPC)  to adopt multi-core
processors and accelerator-based platforms.  
Typical HPC systems today are clusters of computing nodes interconnected by fast
low-latency communication networks, e.g. Infiniband.
Each node typically has two standard multi-core CPUs, 
each attached to one or more accelerators, either Graphic Processing Unit (GPU) 
or many-core systems.

Recent development trends see a common path to performance for CPUs and 
accelerators, based on an increasing number of independent cores and on wider 
vector processing facilities within each core.
In this common landscape, accelerators offer additional computing performance
and better energy efficiency by further pushing the granularity of their data 
paths and using a larger fraction of their transistors for computational data 
paths, as opposed to control and memory structures. 
As a consequence, even if CPUs are more tolerant for intrinsically unstructured
and irregular codes, in both class of processors computing efficiency goes 
through careful exploitation of the parallelism available in the target
applications combined with a regular and (almost) branch-free scheduling of
operations.

This remark supports our attempt to write just one LQCD code which is not only 
portable, but also efficiency-portable across a large number of state-of-the-art 
CPUs and accelerators.
In this paper we consider Intel multi-core CPUs, and NVIDIA and AMD GPUs, 
commonly used today by many scientific HPC communities. 
\tablename~\ref{tab:architecture} summarizes some key features of the 
systems we have used\cite{kepler,pascal,hawaii,haswell}, that we describe very 
briefly in the following. 

%
\begin{table}
\tbl{
Selected hardware features of some of the processors used in this work: the Xeon-E5 systems are two recent 
multi-core CPUs based on the Haswell and Broadwell architecture, the K80 GPU is 
based on the {\em Kepler} architecture while the P100 GPU adopts the
{\em Pascal} architecture. The FirePro W9100 is an AMD GPU, based on the Hawaii 
architecture.
}{
\centering
\resizebox{\textwidth}{!}{
\begin{tabular}{lrrrlrr}
\toprule
                                &  Xeon E5-2630 v3 & Xeon E5-2697 v4 & K80-GK210 &                          & P100        & FirePro W9100 \\  
\midrule                        
Year                            &  2014            & 2016            & 2014      &                          & 2016        & 2014   \\
Architetcure                    &  Haswell         & Broadwell       & Kepler    &                          & Pascal      & Hawaii \\
\#physical-cores / SMs          &  8               & 18              & 13        & \hspace{-1em} $\times$ 2 & 56          & 44     \\
\#logical-cores / CUDA-cores    &  16              & 26              & 2496      & \hspace{-1em} $\times$ 2 & 3584        & 2816   \\    
Nominal Clock (GHz)             &  2.4             & 2.3             & 562       &                          & 1328        &  930   \\
Nominal DP performance (Gflops) &  $\approx 300$   & $\approx 650$   & 935       & \hspace{-1em} $\times$ 2 & 4759        & 2620   \\
\midrule     
LL cache (MB)                   &  20              & 45              & 1.68      &                          & 4           & 1.00   \\
Total memory supported (GB)     &  768             & 1540            & 12        & \hspace{-1em} $\times$ 2 & 16          & 16     \\
Peak mem. BW (ECC-off) (GB/s)   &   69             & 76.8            & 240       & \hspace{-1em} $\times$ 2 & 732         & 320    \\ 
\bottomrule  
\end{tabular}
}
\label{tab:architecture}
}
\end{table}
%

Intel Xeon-E5 architectures are conventional x86 multi-core architectures. We
have used two generations of these processors, differing for the number of
cores and for the amount of integrated last-level cache. Performances in both
cases rely on the ability of the application to run on all cores and to use
256-bit vector instructions.  

NVIDIA GPUs are also multi-core processors. A GPU hosts several Streaming
Multiprocessors (SM), which in turn include several (depending on the specific
architecture) compute units called CUDA-cores.
At each clock-cycle SMs execute multiple warps, i.e. groups of 32 instructions,
belonging to different CUDA-threads, which are executed in {\em Single
Instructions Multiple Threads} (SIMT) fashion.  SIMT is similar to SIMD
execution but more flexible, e.g. different CUDA-threads of a SIMT-group are
allowed to take different branches of the code, although at a performance
penalty. Each CUDA-thread has access to its copy of registers and context
switches are almost at zero cost.
This structure has remained stable across several generations with minor
improvements.
The NVIDIA K80 has two GK210 GPUs; each GPU has 13 {\em Next Generation}
Streaming Multiprocessor, (SMX) running at a base frequency of $562$ MHz that
can be increased to $875$ MHz under specific condition of work-load and power.
The corresponding aggregate peak performance of the two GK210 units is then
$1.87$ and $2.91$ TFlops in double precision. The peak memory bandwidth is
$240$~GB/s considerably higher compared to that of E5-Xeon CPUs.
The GP100 GPU, based on the Pascal architecture, has recently become available.
It has 56 streaming processors running at base-frequency of $1.3$ that can be
increased to $1.48$~GHz, delivering a peak double-precision performance of
$4.76$ and $5.30$~Tflops.  Peak memory bandwidth has been increased to
$732$~GB/s.

The AMD GPUs are conceptually similar to NVIDIA GPU.  The AMD FirePro W9100 has
44 processing units, each one with 64 compute units (stream processors),
running at $930$~MHz.  This board delivers a peak double-precision performance
of $2.6$~Tflops, and has a peak memory-bandwidth of $320$~GB/s.


Native programming models, commonly used for the systems shown 
in \tablename~\ref{tab:architecture}, differ in several aspects. 


For Xeon-E5 CPUs, the most common models are OpenMP and OpenMPI. 
Both models support core-parallelism, running one thread or one MPI 
process per logical core. 
Moreover OpenMP is a directive based programming model, and allows to exploit 
vector-parallelism properly annotating for-loops that can be 
parallelized\cite{openmp}.


On GPUs, the native programming model is strongly based on data-parallel
models, with one thread typically processing one element of the application
data domain.
This helps exploit all available parallelism of the algorithm and hide
latencies by switching among threads waiting for data coming from memory and
threads ready to run.
The native language is CUDA-C for NVIDIA GPUs and OpenCL for AMD systems.
Both languages have a very similar programming model but use a slight different 
terminology; for instance, on OpenCL the CUDA-thread is called work-item, 
the CUDA-block work-group, and the CUDA-kernel is a device program.
A CUDA-C or OpenCL program consists of one or more functions that run either on
the host, a standard CPU, or on a GPU. 
Functions that exhibits no (or limited) parallelism run on the host, while
those exhibiting a large degree of data parallelism can go onto the GPU. 
The program is a modified C (or C++, Fortran) program including keyword
extensions defining data parallel functions, called {\em kernels} or {\em
device programs}.
Kernel functions typically translate into a large number of threads, i.e. 
a large number of independent operations processing independent data items. 
Threads are grouped into blocks which in turn form the execution {\em grid}. 
%
%
When all threads of a kernel complete their execution, the corresponding 
grid terminates. Since threads run in parallel with host CPU threads, 
it is possible to overlap in time processing on the host and the accelerator.


New programming approaches are now emerging, mainly based on directives, 
moving the coding abstraction layer at an higher lever, over the hardware 
details.
These approaches should make code development easier on heterogeneous 
computing systems\cite{openacc}, simplifying the porting of existing 
codes on different architectures.
OpenACC is one such programming models, increasingly used by several scientific 
communities. 
OpenACC is based on \textit{pragma} directives that help the compiler 
to identify those parts of the code that can be implemented as 
{\em parallel functions} and offloaded on the accelerator or divided among 
CPU cores.
The actual construction of the parallel code is left to the compiler making, 
at least in principle, the same code portable without modifications across 
different architectures and possibly offering more opportunities for performance 
portability.
This make OpenACC more descriptive compared to CUDA and OpenCL which are more 
prescriptive oriented.

%
\begin{figure}[t]
\begin{lstlisting}[language=C,label=lst:saxpy,belowcaptionskip=2em,
caption=Sample OpenACC code computing a {\em saxpy} function on vectors 
$x$ and $y$. The {\em pragma} clauses control data transfers between 
host and accelerator and identify the code regions to be run on the 
accelerator.\vspace*{1em}]
 #pragma acc data copyin(x), copy(y) {
 
   #pragma acc kernels present(x) present(y) async(1) 
   #pragma acc loop vector(256) 
   for (int i = 0; i < N; ++i)
     y[i] = a*x[i] + y[i];

   #pragma wait(1); 
 }
\end{lstlisting}
\end{figure}
%

Listing~\ref{lst:saxpy} shows an example of the \textit{saxpy}
operation of the {\em Basic Linear Algebra Subprogram} (BLAS) set coded in 
OpenACC.
The \textit{pragma acc kernels} clause 
identifies the code fragment running on the accelerator,
while \textit{pragma acc loop...} specifies that the iterations of 
the for-loop can execute in parallel.

The standard defines several directives, allowing a fine 
tuning of applications.
As an example, the number of threads launched by each device function and
their grouping can be tuned by the \textit{vector}, \textit{worker} and
\textit{gang} directives, in a similar fashion as setting the number of
\textit{work-items} and \textit{work-groups} in CUDA.
Data transfers between host and device memories are automatically 
generated, and occur on entering and exiting the annotated code 
regions.
Several data directives are available to allow the programmer to
optimize data transfers, e.g. overlapping transfers and computation.
For example, in Listing~\ref{lst:saxpy} the clause \textit{copyin(ptr)} copies
the array pointed by \textit{ptr} from the host memory into the accelerator
memory before entering the following code region; while \textit{copy(ptr)}
perform the additional operation of copying it also back to the host memory
after leaving the code region.
An asynchronous directive \textit{async} is also available, instructing the
compiler to generate asynchronous data transfers or device function executions;
a corresponding clause (i.e. \textit{\#pragma wait(queue)}) allows to wait for 
completion. 

OpenACC is similar to the OpenMP (Open Multi-Processing) framework widely used
to manage parallel codes on multi-core CPUs in several ways\cite{Wienke2014812}; 
both frameworks are directive based, but OpenACC targets accelerators in general, 
while at this stage OpenMP targets mainly multi-core CPUs; the latest release of 
OpenMP4 standard has introduced directives to manage also accelerators, but currently, 
compilers support is still limited.
Regular C/C++ or Fortran code, already developed and tested on traditional CPU
architectures, can be annotated with OpenACC pragma directives 
(e.g. \textit{parallel} or \textit{kernels} clauses) to instruct the compiler to
transform loop iterations into distinct threads, belonging to one or more
functions to run on an accelerator.
Ultimately, OpenACC is particularly well suited for developing scientific HPC 
codes for several reasons:
\begin{itemize}
\item it is highly hardware agnostic, allowing to target several architectures, 
GPUs and CPUs, allowing to develop and maintain one single code version;
\item the programming overhead to offload code regions to accelerators is 
limited to few \textit{pragma} lines, in contrast to CUDA and in particular 
OpenCL verbosity;
\item the code annotated with OpenACC \textit{pragmas} can be still compiled 
and run as plain C code, ignoring the \textit{pragma} directives.
\end{itemize}

\section{OpenACC implementation of Lattice QCD}\label{implementation}

In this section we describe the OpenACC implementation of our LQCD code. 
We first describe the data structures used, then we highlight the most
important OpenACC-related details of our implementation.
%
In writing the OpenACC version, we started from our previous code 
implementations\cite{se4hpcs15}: 
a C++/CUDA\cite{cudacode} developed for NVIDIA GPUs aggressively optimized 
with CUDA-specific features, and a C++ one, developed using OpenMP and MPI 
directives, targeting large CPU clusters\cite{nissa}.

\subsection{Memory allocation and data structures}
\label{memalloc_section}

Data structures have a strong impact on performance\cite{se4hpcs15,ppam15} 
and can hardly be changed on an existing implementation: 
their design is in fact a critical step in the implementation of a new code. 
We have analyzed in depth the impact of data-structures for LQCD on different 
architectures (i.e. a GPU and a couple of CPUs), confirming that the 
{\em Structure of Arrays} (SoA) memory data layout is preferred when using 
GPUs, but also when using modern CPUs\cite{se4hpcs15}.
This is due to the fact that the SoA format allows vector units to process many 
sites of the application domain (the lattice, in our case) in parallel,
favoring architectures with long vector units (e.g. with wide SIMD 
instructions).
Modern CPUs tend indeed to have longer vector units than older ones and we
expect this trend to continue in the future.
%
%
For this reason, all data structures related to lattice sites in our code 
follow the SoA paradigm.

\begin{figure}
  \centering
  \includegraphics[width=0.8\textwidth]{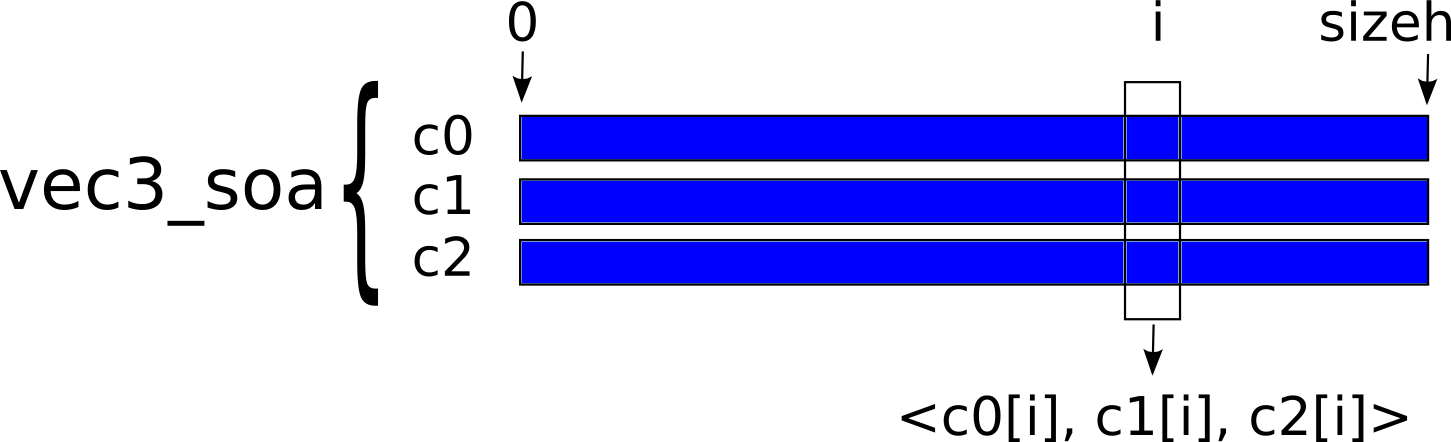}
  \caption{Memory data layout for structure {\sf vec3\_soa}. 
  Each component $i$ of each array {\sf c0, c1} and {\sf c2} is a C99 complex value.  
  See sections~\ref{memalloc_section} for details.}
  \label{fig:mem-vec}
\end{figure}

\begin{figure}
  \centering
  \includegraphics[width=0.6\textwidth]{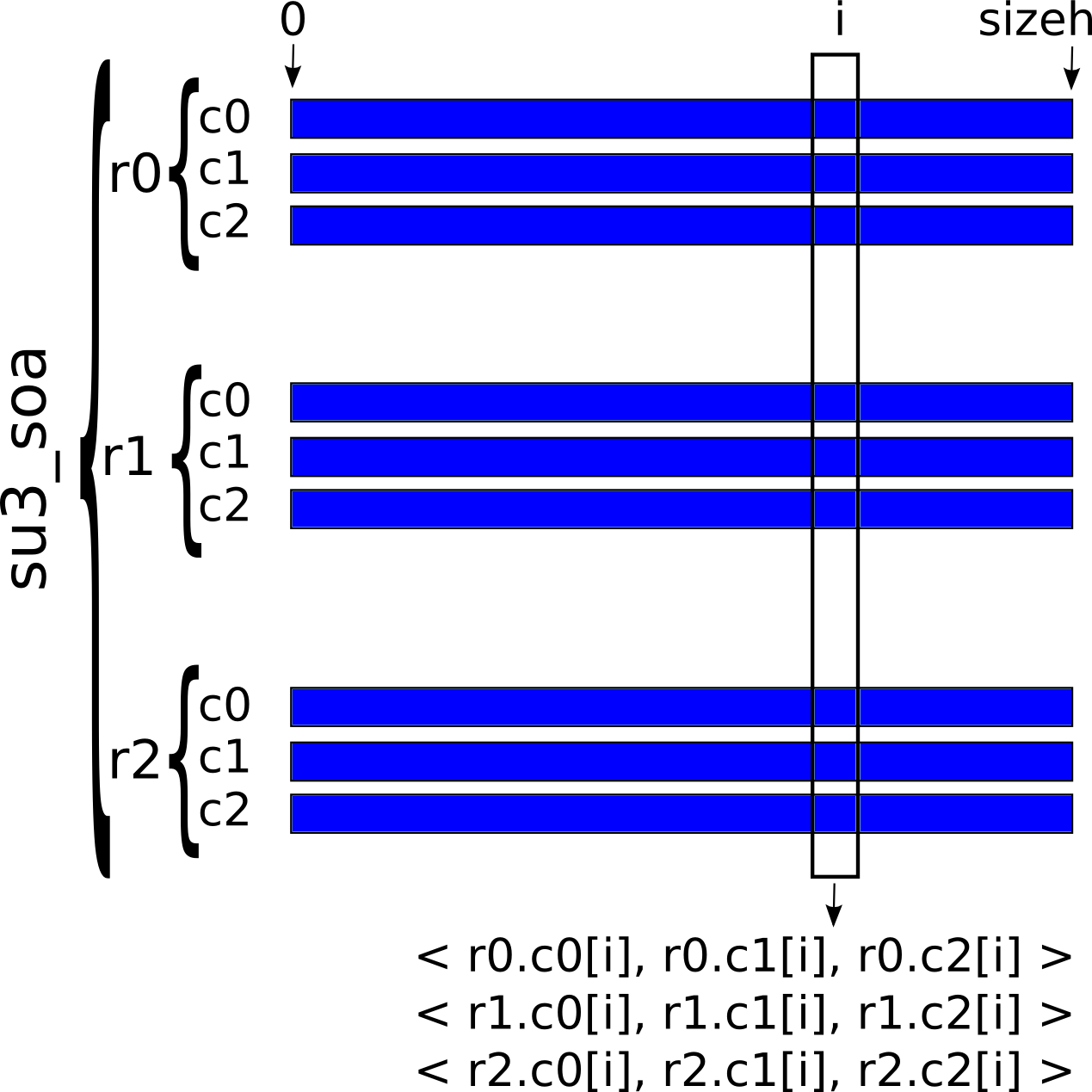}
  \caption{Memory data layout for structure {\sf su3\_soa}, used in 
  the code for $SU(3)$ matrices; this structure contains 3 vectors. 
  To mitigate memory-bandwidth requirements, one can avoid reading 
  and writing the {\sf r2} member and recompute it on the fly, 
  exploiting the unitarity constraint.}\label{fig:mem-su3}
\end{figure}

In our implementation, we use the {\sf C99 double complex} as basic data-type
which allows to use built-in complex operators of the C library making coding
easier and more readable without loss of performance.

The algorithm is based on even/odd preconditioning, so the pseudo-fermion
variables (implemented as {\sf vec3\_soa} data-types) live only on the even
sites of the lattice. 
This comes at the price of requiring that all sides of the lattice must be
even\footnote{Actually, for staggered fermions, this is a requirement coming
from the discretization itself.}; in the following we call {\sf LNH\_SIZEH}
half the number of lattice sites. 
The pseudofermion field has three complex values for each even lattice site, 
corresponding to the three QCD ``colors'' that we label {\sf c0}, {\sf c1}, 
{\sf c2}.
A schematic representation of the {\sf vec3\_soa} structure is shown in 
Fig.~\ref{fig:mem-vec} and a lexicographical ordering was used for the even 
lattice sites: 
\begin{equation}
    \label{lexicographic_eo}
    \textrm{idxh} = \textrm{ \sf (int)} \frac{ x_0 + \textrm{\sf LNH\_N0} 
    [x_1 + \textrm{\sf LNH\_N1} (x_2 + \textrm{\sf LNH\_N2}\, x_3 )]}{2} \qquad 
    \mathrm{s.t.}\ \sum_{i=0}^3 x_i \% 2 =0\ ,
\end{equation}
where {\sf LNH\_N0}, {\sf LNH\_N1} and {\sf LNH\_N2} are the lattice sizes; we
allow for full freedom in the mapping of the physical directions $x$,
$y$, $z$ and $t$ onto the logical directions $x_0$, $x_1$, $x_2$ and $x_3$, as
this option will be important for future versions of the code able to run on
many processors and accelerators. 

The data structure used for the generic $3\times 3$ complex matrices is the
{\sf su3\_soa} data-type,\footnote{Here the name of the data-type is slightly
misleading, since this data structure is used to store $GL(3)$ matrices, while
actual $SU(3)$ matrices require in principle less memory.} used e.g. for
the ``staples'' $\Sigma_{\mu}^{(k)}$ and the $\Gamma_\mu$ matrices needed in the
stouting procedure\cite{Morningstar:2003gk}.  Structure {\sf su3\_soa} is
a collection of 3 {\sf vec3\_soa} structures ({\sf r0}, {\sf r1},
{\sf r2}, see Fig.~\ref{fig:mem-su3}), and data that has to be stored in this
structure typically involve a number of matrices equal to the number of links
present in the lattice, i.e. {\sf 8 LNH\_SIZEH}; this means that an array of 8
{\sf su3\_soa } elements is required.  

Gauge configurations, i.e. the set of the gauge links $U_{\mu}(n)$ and their
stouted counterparts, are stored in memory as an array of 8 {\sf su3\_soa}
structures. As previously explained the algorithm is typically bandwidth
limited and for $SU(3)$ matrices it is convenient to read and write just the
first two rows, computing the third one on the fly as ${\sf r2}=({\sf}r0 \wedge
{\sf r1})^*$. Note that the SoA memory layout avoids the prefetching problems
discussed in similar cases\cite{JooPhi}.

Other data structures are needed to store in memory $3\times 3$ traceless Hermitian
matrices or $3\times 3$ traceless anti-Hermitian matrices. In these
cases, only 8 real parameters per matrix are needed: 3 complex numbers for the
upper triangular part and the first two elements of the diagonal, which are
real (imaginary) numbers for (anti-)Hermitian traceless matrices. These data
structures have been implemented according to the SoA scheme as follows: {\sf
thmat\_soa} and {\sf tamat\_soa} contain 3 vectors of {\sf C99 double complex}
numbers and 2 vectors of {\sf double} numbers, in a form that closely resemble
the one of {\sf vec3\_soa}.

Data movements between device and host are negligible, with significant
transfers happening only at the beginning and at the end of each Monte Carlo
update, and managed mainly with the {\sf update device} and {\sf update host}
OpenACC directives.

\subsection{Implementation methodology}
\label{sec:meth}

To initially assess the performance level achievable using OpenACC, we have developed 
a mini-application benchmark of the Dirac operator\cite{se4hpcs15}.  
As previously underlined this is the fundamental building block of the Krylov
solver,  commonly accounting for not less than $40\%$ of the running time, and
reaching up to $80\%$ in low temperature simulations.  This compute intensive
part of an LQCD  simulation is where most of the optimization efforts are
usually concentrated\cite{dirac-opt}.
The Dirac operator code uses three functions: {\sf deo}, {\sf doe} 
(corresponding respectively to the application of functions $D_{eo}$ and
$D_{oe}$ defined in \Eqref{eq:diracblock}) and a \textit{zaxpy}-like function
which is negligible in terms of execution time.  
A direct comparison indicated that the performance of the OpenACC versions of 
the double precision {\sf deo} and {\sf doe} functions were comparable with 
the CUDA ones\cite{se4hpcs15}.
This promising start was a strong indication that also for LQCD the higher 
portability of the OpenACC implementation is not associated with a serious 
loss of performance, and motivated us to proceed to an OpenACC implementation 
of the full RHMC code.
As a side benefit, the use of the OpenACC programming model significantly 
simplified the implementation of algorithmic improvements.

The implementation of these new features started with the coding and testing of
the improvements on a single thread version. After the algorithm is validated,
the acceleration is switched on by annotating the code with {\sf \#pragma}
directives. 
%
In order to have a more readable code, the most complex kernels have been split
in several functions. While small functions can be used in kernels if declared
as {\sf static inline}, for larger ones we had to use the {\sf routine seq}
OpenACC directive as large functions cannot be inlined.

Kernels have been parallelized following two different approaches. Those using
data  belonging to nearest (and/or next-to-nearest) neighbors have been
parallelized via  the {\sf \#pragma acc loop} directive on $4$ nested loops, one
for each dimension. 
This allows to use 3D thread blocks, which should improve data reuse between
threads  thus reducing bandwidth requirements, which is our major performance
concern. The other kernels, i.e. the ones performing only single-site
operations, have been parallelized using a single cycle running on the
lattice sites.

\begin{table}[ht]
\tbl{
Breakup of the execution time of a selection of computationally heavy steps of 
our OpenACC code on different architectures for low temperature and finite 
temperature simulations.
}{
 \begin{tabular}{l|S[table-format={2}]
                   S[table-format={2}]
                   S[table-format={2}]
                   S[table-format={2}]}
 \toprule
 \multirow{2}{*}{Phase} & \multicolumn{2}{c}{GPU NVIDIA GK201} & \multicolumn{2}{c}{CPU Intel E5-2630v3}\\
                        & {Low Temp.} & {High Temp.}  & {Low Temp.} & {High Temp.}\\ 
 \midrule
 Dirac Operator         & 63          & 16            & 57          & 24  \\
 Gauge MD               & 8           & 56            &  1          & 24  \\ 
 \bottomrule
 \end{tabular}
\label{tab:functions}
}
\end{table}

After the implementation of a first full working OpenACC simulation, various
optimization iterations took place, in particular for the performance critical
steps.
These include the Dirac operator in the first place, but also the gauge part of
the molecular dynamics steps, since their relative impact on the overall 
execution time is very large, as shown in \tablename~\ref{tab:functions} for a 
few representative examples.

During the full development phase, every time a new OpenACC feature has been
introduced, extensive checks have been performed to ensure the correctness of
the improved code, against possible semantic misunderstanding of OpenACC
clauses or compiler bugs.

%
%


\subsection{Implementation details of selected kernels}
\label{sec:kernels}

This section describes the overall structure of our code, and focuses 
on the OpenACC implementation of selected performance-critical parts.

%
\begin{algorithm}[ht]
\caption{Top level scheme of the full simulation code}
\label{big_scheme_of_things_algorithm}
\begin{algorithmic}[1]
\STATE Read gauge configuration $U$
\STATE Create momenta $p$
\STATE Generate pseudofermions by heatbath \label{pseudoferm_generation_algorithm}
\STATE Calculation of initial action
\STATE Molecular Dynamics [possibly in single precision] \label{moldyn_step_algorithm}
\STATE Calculate action variation $\Delta S$ \label{final_action_calculation_algorithm}
\STATE Montecarlo step accepted with probability $\textrm{min}(1,e^{-\Delta S})$
\STATE Take measurements
\end{algorithmic}
\end{algorithm}
%

Algorithm~\ref{big_scheme_of_things_algorithm} is a top-level description 
of the full code, showing the main computational tasks. 
For performances, the most critical steps are Molecular Dynamics 
(step~\ref{moldyn_step_algorithm}) followed by the heatbath generation of 
the pseudofermions (step~\ref{pseudoferm_generation_algorithm}), 
and the calculation of the final action (step~\ref{final_action_calculation_algorithm}).
Steps~\ref{pseudoferm_generation_algorithm}
and~\ref{final_action_calculation_algorithm}  consist basically in function
calls to the multishift inverter routine, with a high target accuracy.


The outer level of the multistep integrator for Molecular Dynamic evolution 
(step~\ref{moldyn_step_algorithm}) in Algorithm~\ref{big_scheme_of_things_algorithm}
is expanded in Algorithm~\ref{outer_cycle_algorithm}. 
As explained in Sec.(\ref{grand_challenge_section}), in zero temperature 
simulations or for small quark masses usually the heaviest computational parts
are the calculations of the fermion force, while in high temperature simulations 
the load is shifted inside the gauge cycles, as already shown 
in \tablename~\ref{tab:functions}. 
The fermion force calculation step is implemented
following\cite{Morningstar:2003gk}; for this step a large fraction of the
execution time is is spent in computation of {\sf deo} and {\sf doe} functions
implementing the Dirac operator.


\begin{figure}[ht]
\begin{lstlisting}[language=C,label=lst:dirac,belowcaptionskip=2em,
caption={OpenACC implementation of the {\sf Deo} function; directive 
{\sf vector tile} divides the computational domain in sub-lattices (tiles), 
each processed within a compute unit in order to allow data re-use.\vspace*{1em}}]
void acc_Deo( __restrict const su3_soa * const u,
                   __restrict vec3_soa * const out,
                   __restrict const vec3_soa * const in,
                   __restrict const double_soa * const backfield){
  int hd0, d1, d2, d3;
  #pragma acc kernels present(in) present(out) 
                      present(u)  present(backfield) async(1)
  #pragma acc loop independent gang(GANG)
  for(d3=0; d3<nd3;d3++) {
    #pragma acc loop independent vector tile(TILE0,TILE1,TILE2)
    for(d2=0; d2<nd2; d2++) {
      for(d1=0; d1<nd1; d1++) {
        for(hd0=0; hd0 < nd0h; hd0++) {
          ...        
        }
      }
    }
  }
}
\end{lstlisting}
\end{figure}


The {\sf deo} OpenACC implementation is shown in Listing~\ref{lst:dirac},
showing the 4 dimension nested loops and the corresponding pragma directives.
In this listing OpenACC directives are used: i) to identify the data structures
already present in the accelerator memory, when targeting accelerators ({\sf
present()} clause); ii) to make the compiler aware of the data independence of
loops iterations ({\sf independent} clause); iii) to request to group
iterations in order to execute them in the same (or close) compute units  ({\sf
tile} clause).
In particular, the {\sf tile} OpenACC clause asks the compiler to split or
strip-mine each loop in the nest into two loops, an outer tile loop and an inner
element loop.

Where possible (e.g. in {\sf deo} and {\sf doe}), performing computations of
adjacent lattice sites in close hardware compute units may increase data reuse
(i.e. matrices shared between sites)\cite{dirac-opt} for all the architectures
where data caches are present, which means almost every modern processing
architecture.
The tile sizes offering the best performance depend, for each kernel, on
several features of each specific architecture, e.g. vector units size,
register numbers, cache levels and sizes.
We keep a door open for limited architecture-specific optimization, by
allowing to specify the {\sf TILE0, TILE1, TILE2} variables at 
compile time, telling the compiler how to group together iterations involving 
adjacent lattice sites.


%
\begin{algorithm}[t]
    \caption{MD evolution - 2nd order Minimum Norm integrator (outer cycle)}
    \label{outer_cycle_algorithm}
\begin{algorithmic}[1]
    \STATE Fermion Force Calculation 
    \STATE Evolve momenta for $\lambda \Delta T/N_{md}$  \, \, \, \,  \, \,  \, \,  \, \,  \, \, \, \, \, \,  \COMMENT{$\lambda=0.1931833275037836$}\cite{Omelyan03}
    \FOR{$i=1$ to $N_{md}-1$}
\STATE {\bf Gauge cycle } ($\Delta T/2N_{md}$)
\STATE Fermion Force Calculation
  \STATE Evolve momenta for $(1-2\lambda )\Delta T/N_{md}$
  \STATE {\bf Gauge cycle }($\Delta T/2N_{md}$)
\STATE Fermion Force Calculation
  \STATE Evolve momenta for $2\lambda \Delta T/N_{md}$
\ENDFOR
\STATE {\bf Gauge cycle }($\Delta T/2N_{md}$)
\STATE Fermion Force Calculation
\STATE Evolve momenta for $(1-2\lambda )\Delta T/N_{md}$
\STATE {\bf Gauge cycle }($\Delta T/2N_{md}$)
\STATE Fermion Force Calculation
\STATE Evolve momenta for $\lambda \Delta T/2N_{md}$
\end{algorithmic}
\end{algorithm}
%

The actual evolution of the gauge configuration happens inside the inner gauge 
cycles, where the gauge contribution to the momenta evolution is also 
calculated. 
Among the tasks performed in the gauge cycles, the computation of staples in the
gauge force calculation is the most time consuming. It consists of calculating 6
products of 3 and 5 $SU(3)$ matrices representing links on C-shaped paths on the
lattice. 

The implementation of one of these functions is sketched in Listing~\ref{lst:rect_staples}: 
also in this case the parallelization has been done using the 
{\sf tile} directive over the 3 innermost nested cycles. 
This allows us also in this case to use 3D thread blocks, which should improve 
data reuse between threads, reducing the bandwidth needs.
We shall also remark that in this case, since second-nearest-neighbor-site
addressing is needed, for the sake of simplicity we use indirect
addressing\footnote{The code would be greatly more complicated if using direct
addressing, also because of some limitations in the coding options necessary to
avoid branches that would destroy thread coherence.}. 
Notice that the function {\sf staple\_type1} (as well as similar ones) has to be 
declared with {\sf \#pragma acc routine seq} to be used inside a kernel. 

%
\begin{figure}
\begin{lstlisting}[language=C,label=lst:rect_staples,belowcaptionskip=1em,
caption={Implementation of the function performing the evaluation of a staple.\vspace*{1em}}]
 #pragma acc routine seq
 void staple_type1(...){...}

 void calc_staples_type1( __restrict const su3_soa * const u,
                          __restrict su3_soa * const loc_stap ) {
   int d0, d1, d2, d3, mu, iter;
   #pragma acc kernels present(u) present(loc_stap) 
                       present(nnp_openacc) present(nnm_openacc)
   #pragma acc loop independent gang(IMPSTAPGANG3)
   for(d3=0; d3<nd3; d3++){
     #pragma acc loop independent vector tile(IMPSTAPTILE0,
                                              IMPSTAPTILE1,
                                              IMPSTAPTILE2)
     for(d2=0; d2<nd2; d2++) {
       for(d1=0; d1<nd1; d1++) {
         for(d0=0; d0 < nd0; d0++) {
           #pragma acc loop seq
           for(mu=0; mu<4; mu++){
             ...
             const int idx_pmu = nnp_openacc[idxh][mu][parity];
             ...
             staple_type1(&u[dir_nu_1R], idx_pmu, ... )
\end{lstlisting}
\end{figure}
%

In order to improve performance, we also implemented a single precision version
of the code for the molecular dynamics evolution. Due to the low arithmetic
density of the LQCD algorithms, on GPUs at least, all kernels are memory-bound;
this means that, when precision is not an issue, it is preferable to have
single precision versions of selected functions and structures, as a plain
$\times2$ increase in performance is expected with respect to the double
precision implementation.

\section{Performance analysis}\label{results}

To compare the performance of our code on different architectures we consider 
two different benchmarks taking into account the most computational 
intensive parts of the code.
The first benchmark evaluates the performance of the Dirac operator, 
both single and double precision version, and the latter evaluates 
the performance of the gauge part of the molecular dynamics step.
Depending on input configuration parameters either the former or the latter
kernels make up most of the execution time of a typical simulation, as shown in
\tablename~\ref{tab:functions}.  

We present the execution time per site of the Dirac operator for different 
lattice sizes in \tablename~\ref{tab:dirac-norm}.
Exactly the same code has been run on all platforms without requiring any change; 
we have just re-compiled it with different flags instructing the PGI 16.10 
compiler to target the corresponding architectures and using the best tile
dimensions for each of them.

\begin{table}[ht]
\tbl{Measured execution time per lattice site [ns] for the Dirac operator, 
on several processors and for several typical lattice sizes.}{
\resizebox{\textwidth}{!}{
\begin{tabular}{l|SSSSSSSS}
\toprule
\multirow{3}{*}{Lattice}   & \multicolumn{8}{c}{Processor (CPU or GPU)} \\
                           & \multicolumn{2}{c}{NVIDIA GK201} 
                           & \multicolumn{2}{c}{NVIDIA P100}  
                           & \multicolumn{2}{c}{Intel E5-2630v3} 
                           & \multicolumn{2}{c}{Intel E5-2697v4} \\
                           & { SP} & { DP} & { SP} & { DP} & { SP} & { DP} & { SP} & { DP} \\
\midrule
\midrule
 $32^2 \times 8 \times 32$ & 4.19 &  8.51 & 1.77 & 3.07 & 72.18 &  99.17 & 38.92 &  54.90 \\
 $32^3 \times 8$           & 4.15 &  8.39 & 1.22 & 2.48 & 72.81 & 101.46 & 77.33 & 103.87 \\
 $24^4$                    & 4.43 &  8.62 & 1.58 & 2.90 & 70.44 &  94.42 & 51.13 &  66.87 \\
 $32^4$                    & 4.02 &  9.54 & 1.32 & 2.40 & 79.05 & 100.19 & 43.90 &  54.88 \\
 $32^3 \times 36$          & 4.03 &  8.48 & 1.46 & 2.54 & 83.12 & 107.47 & 38.82 &  50.29 \\
\bottomrule
\end{tabular}
}
\label{tab:dirac-norm}
}
\end{table}



We tested two different NVIDIA GPUs, the K80 based on the Kepler architecture 
and the recently released P100 board based on the Pascal architecture.
For the K80 the single precision version takes $\approx 4ns$ per lattice site, 
while the double precision version requires $\approx 8.5ns$.
Running on the P100 we measure $\approx 1.5ns$ for single and $\approx 2.5ns$ 
for double precision, improving approximately by a factor $3 \times$ over the K80. 
This results perfectly scales with architecture performance of P100 that  
has $\approx 4.3 \times$ more cores and $\approx 3 \times$ more memory bandwidth, 
see \tablename~\ref{tab:architecture}.


Concerning Intel CPUs, we have compared two different processors, the 8-core 
E5-2630v3 CPU based on Haswell architecture, and the 18-core E5-2697v4 CPU based
on Broadwell.
Since computing resource of the CPUs are roughly $3 \times$ lower than on GPUs,
see \tablename~\ref{tab:architecture}, a performance drop is expected. 
However, the actual performance drop measured on both CPUs is much larger than
this expected theoretical figure; indeed time per site is approximately $10
\times$ or larger on the Haswell than on one K80 GPU.  The Broadwell performs
approximately a factor $2\times$ better compared to Haswell, at least for some
lattice sizes.
We have identified two main reasons for this non-optimal behavior, and both of
them point to some still immature features of the PGI compiler when targeting
x86 architectures, that -- we expect -- should be soon resolved:

\begin{itemize}

\item \textbf{Parallelization} - the compiler is only able to split outer-loops
across different threads, while inner loops are executed serially or vectorized
within each thread.
This explains why on the Broadwell CPU running on a lattice
$32^2\times8\times32$ we have a performance $2\times$ better than for a
$32^3\times8$ lattice, which has the same volume but allows to split the outer
loop only on $8$ threads.
 
\item \textbf{Vectorization} - as reported by the compilation logs, the
compiler fails to vectorize the {\sf deo} and {\sf doe} functions computing the
Dirac operator (see Listing~\ref{lst:dirac}) reporting to be unable to
vectorize due to the use of ``mixed data-types''.
To verify if this is related to how we have coded these functions, we have
translated the OpenACC pragmas  into the corresponding OpenMP ones -- without
changing the C code -- and  compiled using the Intel compiler (version 17.0.1).
In this case the compiler succeeds in vectorizing the two functions, running a
factor $2 \times$ faster compared to the OpenACC version compiled by PGI
compiler.
\end{itemize}

%
\begin{table}[ht]
\tbl{
Measured execution time per lattice site [ns] for the pure gauge Molecular Dynamics 
step for several processors and several typical lattice sizes.
}{
\resizebox{\textwidth}{!}{
\begin{tabular}{l|SSSS}
\toprule
\multirow{2}{*}{Lattice} & \multicolumn{4}{c}{Processor (CPU or GPU)} \\
                         & \multicolumn{1}{c}{NVIDIA GK201}
                         & \multicolumn{1}{c}{NVIDIA P100}
                         & \multicolumn{1}{c}{Intel E5-2630v3}
                         & \multicolumn{1}{c}{Intel E5-2697v4}\\
\midrule
\midrule
 $32^2 \times 8 \times 32$ & 193.79 & 51.80 & 1613.88 &  926.48 \\
 $32^3 \times 8$           & 190.39 & 51.69 & 2075.08 & 1756.78 \\
 $24^4$                    & 212.04 & 53.74 & 1265.13 &  979.28 \\
 $32^4$                    & 201.82 & 51.72 & 1719.97 &  944.40 \\
 $32^3 \times 36$          & 208.54 & 52.62 & 1801.81 &  837.68 \\
\bottomrule
\end{tabular}
}
\label{tab:md-norm}
}
\end{table}
%

\tablename~\ref{tab:md-norm} shows the execution time of the gauge part of the
molecular dynamics step. As already remarked this is one of the two most
time-consuming steps together with the application of the Dirac operator.  As
we see the update time per site is quite stable for all lattice sizes we have
tried and for all architectures. Going from the NVIDIA K80 to the P100 the time
improves by a factor $\approx 3\times$, while between Haswell and Broadwell we
have roughly a factor $\approx 1.5\times$ / $2.0\times$.

We finally mention that we have also been able to compile and run our code on an
AMD FirePro W9100 GPU and on the latest version of the Intel Xeon Phi processor,
the Knights Landing (KNL). However, in these cases, results are still
preliminary. In more details, the compiler itself crashes when compiling the
code for the AMD GPU for some specific lattice sizes; for the KNL, specific
compiler support is still missing, but this processor is able to run the code
compiled for the Haswell architecture, implying however that 512-bit
vectorization is not used. These problems do not allow us to perform  a
systematic comparison of performance for these architectures.  Once again, we
believe that  this is due to some immaturity of the compiler, and we expect that
these issues will be resolved in future versions. 




\begin{table}[ht]
\tbl{
Execution time [sec] of a full trajectory of a complete Monte Carlo
simulation for several typical physical parameters, running on one GPU of a 
NVIDIA K80 system. We compare the OpenACC code developed in this paper and 
and earlier GPU-optimized CUDA code. Here we use  
the standard Wilson action and unimproved staggered fermions as the CUDA code
does not support the more advanced improvements available in the OpenACC version.
}{
{                                          
\begin{tabular}{l|SSSSS}
\toprule                                                             
{Lattice}        & $am$   & $\beta$ & {CUDA}  & {OpenAcc} & {Variation} \\
\midrule                           
\midrule            
  $32^3 \times 8$ & 0.0125 & 5.55   &  392.69 &   490.74  & {+25\%}  \\
  $24^4$          & 0.0125 & 5.55   &  303.80 &   328.07  &  {+8\%}  \\
  $32^4$          & 0.001  & 5.52   & 8973.82 &  8228.36  &  {-8\%}  \\
\bottomrule
\end{tabular}
}
\label{tab:cuda-openacc}
}              
\end{table} 

\tablename~\ref{tab:cuda-openacc} addresses the question of the efficiency
costs (if any) of our architecture-portable code; the table compares the
execution time for a {\em full} Monte Carlo step (in double precision) of the
OpenACC code and a previously developed CUDA implementation\cite{cudacode},
optimized for NVIDIA GPUs. Although the two codes are not exactly in a one to
one correspondence, the implementations are similar enough to make such a test
quantitatively meaningful. 
One immediately sees that the performances of the two implementations are
comparable and the use of OpenACC does not imply a dramatic performance loss,
the differences between the execution times of the two versions being of the
order of $10\div 20\%$.

The worst case is the one of the $32^3\times 8$ lattice, in which OpenACC is
about $25\%$ slower than CUDA.  Since we are comparing an high-level version of
the code with one specifically developed for NVIDIA GPUs, this would not be a
dramatic loss, however in this case the comparison is also not completely fair.
Indeed for this high temperature simulation the gauge part of the Molecular
Dynamic step starts to be the computationally heaviest task and, in the CUDA
implementation, part of it had been explicitly hard coded in single precision.

For the low-temperature test cases the differences between the CUDA and the
OpenACC implementation are much smaller and, in fact, in one case the OpenACC
version is the fastest one. A possible explanation of this is the following: in
the CUDA version unidimensional blocks are adopted to parallelize the Dirac
operator, while in the OpenACC implementation three-dimensional block
structures are used, that fit better the larger cache of recent GPUs and,
especially on larger lattices, improves data reuse.

%
%


\section{Concluding Remarks}\label{conclusions}

In this work we have developed a full state-of-the-art production-grade code 
for Lattice QCD simulations with staggered fermions, using 
the OpenACC directive-based programming model. 
Our implementation includes all steps of a complete simulation, and most of 
them run on accelerators, minimizing the transfer of lattice data
to and from the host.
We have used the PGI compiler, which supports the OpenACC standard and is able
to target almost all current architectures relevant for HPC computing, even if
with widely different levels of maturity and reliability.
Exactly the same code runs successfully on NVIDIA many-core GPUs and Intel
multi-core CPUs,  and for both architectures we have measured roughly
comparable levels of efficiency. Also, the performance of the complete code is
roughly the same as that of an equivalent code, specifically optimized for
NVIDIA GPUs and written in the CUDA language.
Our code also runs on AMD GPUs and on the KNL Intel Phi processor, even if the
compilation and run-time environment for these processors is still unable to
deliver production-grade codes; in these cases, we have strong indications that
these problems come from a residual immaturity of the compilation chain and we
expect that they will be soon resolved. 
All in all, our final result is a LQCD Monte Carlo code portable on a large
subset of HPC relevant processor architectures and with consistent
performances.
    
Some further comments are in order: i) using a directive-based programming model, we are able to target different computing platforms presently
used for HPC, avoiding to rewrite the code when moving from one platform to 
another; ii) the OpenACC standard provides a good level of hardware abstraction
requiring the programmer to only specify the function to be parallelized and
executed on the accelerator; the compiler is then able to exploit the
parallelism according to the target processor, hiding from the programmer most
hardware optimizations; iii) the OpenACC code has roughly the same level of
performance of that implemented using a native language such as CUDA for NVIDIA
GPUs, allowing to efficiently exploit the computing resources of the target
processor.

In the near future we plan to carefully assess performances on AMD and KNL
systems, in order to enlarge the platform portfolio of our code.  We also plan
to assess whether OpenMP4 provides the same level of portability as OpenACC, as
soon as compilers supporting this programming standard become available.  This
is important to have a unique directive-based programming model which is
widely used by several scientific communities and supported by several
compilers (GCC, ICC, \ldots).  Finally, we are already working on a massively
parallel version of our code, able to run concurrently on a large clusters of
CPUs and accelerators.


\section*{Acknowledgments}
We warmly thank Francesco Sanfilippo (the developer of the NISSA
code\cite{nissa}) for his advice and support. We thank  the INFN Computing
Center in Pisa for providing us with the development framework, and
Universit\`a degli Studi di Ferrara and INFN-Ferrara for the access to the COKA 
GPU cluster.
This work has been developed in the framework of the SUMA, COKA and COSA
projects of INFN.  FN acknowledges financial support from the INFN SUMA
project.





\end{document}